\documentclass[prb,twocolumn,showpacs,amsmath,amssymb]{revtex4}

\usepackage{graphicx}
\usepackage{dcolumn}
\usepackage{bm}
\begin{document}

\title{Probing multiferroicity and spin-spin interactions via dielectric
measurements on Y-doped HoMnO$_{3}$ in high magnetic fields}
\author{R. Vasic, H. Zhou, E. Jobiliong, C.R.  Wiebe, and J.S. Brooks}
\affiliation{Department of Physics and National High Magnetic
Field Laboratory, Florida State University, Tallahassee, Florida
32310, USA\\}
\date{July 27, 2006}%

\begin{abstract}
The magnetic field dependent phase diagram of the ferroelectric
material ${\rm Ho}_{1-x}Y_{x}MnO_{3}$ has been investigated for x
= 0, 0.6, and 0.8 for fields directed along both the ab in-plane
and c-axis basal plane directions. Dielectric measurements are
used to map out the re-entrant temperature-magnetic field phase
transitions which involve in-plane Mn spin rotations in the
antiferromagnetic state below the N$\acute{e}$el temperature. We
show through the alloy study that the Ho sublattice plays a major
role in all transitions, and that the phase diagram for
$Ho_{1-x}Y_{x}MnO_{3}$ is dependent on magnetic field direction.
We describe this behavior in terms of the interaction of the Ho
sublattice spin system with the underlying, robust $YMnO_{3}$
antiferromagnetic triangular lattice, where the Ho-Y spin
interactions are highly sensitive to concentration and field
direction.
\end{abstract}

\pacs{75.47.Lx,77.80.-e, 77.22.-d, 66.70.+f, 65.40.De}

\maketitle The rare-earth hexagonal materials $RMnO_{3}$(R=Ho, Lu,
Y) have attracted considerable recent attention due to the
uncommon coexistence of coupled ferroelectric and
antiferromagnetic orderings.\cite{fiebig1} Exemplary in the
understanding of this family of compounds is ${YMnO}_{3}$,which
has a ferroelectric transition at $T_{C}=900$~K.\cite{bertaut}
This gives rise to a small distortion of the two dimensional
networks of triangular ${Mn}^{3+}$(S=2) ions out of the basal
plane. In principle, the Mn-Mn spin interactions are highly
frustrated, but this lattice distortion relieves some of the
frustration, and results in an antiferromagnetic ordering at 70~K
with a 120 degree spin structure.\cite{huang} Hence, the
ferroelectric and magnetic order can coexist with a strong
coupling between the two disparate phenomena.\cite{hill}

The compound ${HoMnO}_{3}$, where the nonmagnetic $Y^{3+}$ ions
have been replaced with $Ho^{3+}$, is a more complex system
due to the addition interaction of $Ho^{3+}$ spins.\cite{fiebig1}
Although the ferroelectric temperature $(T_C=875~{\rm K})$ does
not change significantly from $YMnO_3$, there are a series of
lower temperature magnetic transitions.\cite{coeure}  The
$Mn^{3+}$ spins order antiferromagnetically at $T_{N}= 76~K$
\cite{fiebig2} and in Fig. 1, a description of the spin
configurations with decreasing temperature below $T_{N}$ is
presented. Here there is a subsequent spin reorientation
transition at 40~K($T_{\rm SR}$), which is accompanied by a small
moment forming on the Ho site.\cite{vajk, lorenz, munoz, lonkai,
fiebig3} The Ho spins themselves order completely at $T_{Ho}$=5~K,
where a second Mn spin reorientation occurs.\cite{lottermoser} The
complete magnetic structure is still a matter of debate, but
recent measurements have shown that at $T_{N}$, the Mn spins are
in a 120 degree antiferromagnetic structure, at $T_{SR}$ there is
a 90 degree rotation of these spins accompanied by a partial
ordering of the Ho spins along the c-direction, and at $T_{Ho}$,
there is a final reorientation of Mn spins in the basal plane,
with a complete AF ordering of the Ho spins.\cite{fiebig1, vajk}
At temperatures at and below $T_{Ho}$, recent measurements on flux
grown single crystals have shown that additional phases appear as
a function of magnetic field.\cite{fiebig4, yen}

The coupling of the ferroelectric and magnetic order parameters to
the lattice result in a high sensitivity of a number of different
probes, including dielectric effects, to the mapping of the
magnetic phases.\cite{lee, cruz, litvinlorentPRL, katsufuji}  In
the present work, we focus on dielectric measurements to explore
the relative effects of Ho, Y, Mn, and Ga in the alloy series
$Ho_{1-x}Y_{x}MnO_{3}$ and $HoMn_{1-y}Ga_{y}O_{3}$. The emphasis
is on the anisotropic high magnetic field properties of image
furnace grown single crystals in the temperature range $T_N < T <
T_{Ho}$.

\section{\label{sec:level1}EXPERIMENTAL}
Single crystals of  $HoMnO_{3}$, $Ho_{1-x}Y_{x}MnO_{3}$(x=0; 0.6;
0.8) and $HoMn_{1-y}Ga_{y}O_{3}$(y=0.1) were grown by a
traveling-solvent-floating zone (TSFZ) technique\cite{hdzhou}.
Samples with typical dimensions $0.5 \times 0.5 \times 5.0 {\rm
mm}^{3}$ were oriented, cut and polished for dielectric
measurements with parallel plate silver electrodes normal to the c
direction. A standard ac capacitance bridge method was employed
where the rms electric field amplitude applied between plates in c
direction was 50~V/cm, substantially less than the $10^5$ V/cm
field used to stabilize the polarized Ho state.\cite{fiebig3} The
real (capacitative - C) and loss (dissipative - D) signals were
measured at 100 kHz vs. temperature in both low field
superconducting and high field resistive magnets at the National
High Magnetic Field Laboratory. In all cases presented here the ac
electrical field was parallel to the $c$ axis. The magnetic field
anisotropy was measured simultaneously for two samples of each Y
concentration, cut from the same crystal to compare directly
anisotropic effects of magnetic field in ab plane and $c$
direction. To avoid strong torque effects,  care was taken to
affix the samples to avoid movement in high magnetic fields.

\section{\label{sec:level1}RESULTS\protect\\}

The transitions where Mn spin rotations occur can be observed by a
number of different methods, including magnetization, ac magnetic
susceptibility and specific heat\cite{lorenz}, neutron
scattering\cite{lonkai, vajk}, lattice constant\cite{yen}, thermal
conductivity\cite{sharma} and due to the ferroelectric and
magnetic coupling, dielectric
measurements\cite{yen,litvinlorentPRL}. Dielectric studies, the
focus of the present work, are particularly useful for the mapping
of the temperature dependent magnetic phases. Our results for the
two field orientations, B//c and B//ab are shown in Figs. 2 - 4
for $Ho_{1-x}Y_xMnO_{3}$ and in Fig. 5 for
$HoMn_{0.9}Ga_{0.1}O_{3}$.  All data show the signature (peak in
the dielectric constant) and field dependence of the higher
temperature $T_{SR}$ transition, and the lower temperature
$T_{Ho}$ transition is also evident in some cases. (Our data do
not extend to lower temperatures where additional
phases\cite{fiebig4, yen} are observed.) Our measured value of
dielectric constant for all samples was approximately 16, in
agreement with other studies\cite{yen,cruz,litvinlorentPRL}.

    In Fig. 6 (for B//c) and Fig. 7 (for B//ab) we summarize the
field and temperature dependent dielectric peak signatures
observed for all samples and field directions. The most prominent
features are the $T_{SR}$ and $T_{Ho}$ phase boundaries. We focus
first on the $HoMnO_{3}$ (i.e. x = 0) phase diagram in Fig. 6,
where the full width-half maxima of the dielectric peaks have been
used to determine the widths of the transitions. The data show
overall agreement with previous determinations of the B//c
``re-entrant" phase diagram\cite{litvinlorentPRL}. In the Fig. 7,
referring first to the x= 0 data phase diagram for the B//ab-plane
data, we find that the slope $dT_{SR}$/d(B//ab) is significantly
less, that $T_{SR}$ persists to significantly higher fields, and
that the re-entrant character of the phase diagram is nearly
gone.(In Fig. 7 only the peak values for $T_{SR}$ and $T_{Ho}$ are
plotted)

    Turning next to the doping study, we find that for B//c (Fig. 6) $T_{SR}$
increases with increasing Y concentration, and that in general the
re-entrant phase boundaries expand in both their range of
temperature and field. For B//ab, $dT_{SR}$/d(B//ab) decreases
with increasing Y concentration, and for x = 0.6 and 0.8, the
lower transition normally attributed to $T_{Ho}$ vanishes, and the
re-entrant phase boundary is gone. For $HoMn_{0.9}Ga_{0.1}O_{3}$
which represents a small non-magnetic substitution of Ga on the Mn
site, we found that Ga (y=0.1) increases $T_{SR}$ and decreases
$T_{Ho}$( Fig. 7) expanding $P\underline{6_{3}}c\underline{m}$
phase region.

There are some other aspects of the data that are noteworthy. The
widths of the peaks in the phase diagram have been interpreted as
representing an intermediate ``INT" phase by Lorenz et al.
\cite{lorenz}. Not only the widths, but the amplitudes of the
peaks in the dielectric response at the transitions exhibit a
significant dependence on temperature, field, and Y concentration.
In Fig. 8 we summarize the behavior of the the peak heights vs.
their loci in magnetic field and temperature for the different
alloy samples. Most notably, the peak heights are generally
non-monotonic with field and temperature, and show pronounced
maxima for the more concentrated Y samples. The field/temperature
dependence of the $T_{SR}$ transition widths (not plotted, but
evident by examination of the T-B widths of the re-entrant phase
boundaries in Fig. 6), also show a dependence on Y concentration,
temperature, and magnetic field.

\section{\label{sec:level1}DISCUSSION:\protect\\}

The main results of the present work are that: (1) Holmium plays a
major role in all aspects of the low temperature-magnetic phase
diagram of $Ho_{1-x}Y_{x}MnO_{3}$; (2) in-plane magnetic field
greatly affects the influence of Holmium on the spin rotation
transitions; (3) and, for increasing Y concentration $(x\to 1)$ in
$Ho_{1-x}Y_{x}MnO_{3}$, the $T_{SR}$ transition approaches $T_{N}$
and the $P\underline{6_{3}}c\underline{m}$ phase region, which is
the same $P\underline{6_{3}}c\underline{m}$ symmetry that is
observed in pure $YMnO_{3}$ below $T_{N}$ \cite{fiebig3}, expands
in both temperature and magnetic field range.
    Our results, as represented in Figs. 7 and 8, show that for either/or
in-plane magnetic field and reduced Ho concentration, the lower
temperature re-entrant phase boundary is removed. Only in the case
for x = 0 is $T_{Ho}$ evident for B//ab. Below we discuss the
behavior of the dielectric constant at $T_{SR}$, and the
significance of the alloy and field direction studies in turn.

\subsection{\label{sec:level2} Dielectric response at $T_{SR}$}

    The origin of the spin rotation is due to the presence of the Ho spin
sublattice, where the tendency of the Ho to form magnetic order is
accompanied by in-plane adjustments of the original native
$P\underline{6_{3}}$ c\underline{m} symmetry of the pure
$YMnO_{3}$ triangular lattice AFM order below $T_{N}$. The
dielectric measurements show an increase in dielectric constant as
the re-entrant phase boundaries are crossed between the
$P\underline{6_{3}}c\underline{m}$ and
$P\underline{6_{3}}$\underline{c}m symmetries.  Following the
arguments of Lorenz et al. \cite{lorenz} it is expected that a
lower symmetry $P\underline{6_{3}}$ Mn spin configuration is
present as the spins rotate from one configuration to the other.
This intermediate state causes the dielectric constant to
increase; indicating additional spin-lattice strain effects are
present during the transition. However, the dielectric constant is
nearly the same in either of the adjacent
$P\underline{6_{3}}c\underline{m}$ and
$P\underline{6_{3}}$\underline{c}m phases.
    Although theoretical
progress has been made in describing the electric polarization
susceptibility change at $T_{N}$ due to coupling between AFM and
ferroelectric ordering\cite{zhong}, a similar description for the
$T_{SR}$ transition has not yet been treated. It is reasonable to
expect, however, that the change in the dielectric constant is
mainly coupled to the $S_{i}\cdot S_{j}$ term in the
magnetoelectric coupling Hamiltonian\cite{katsufuji}. Here $S_{i}$
and $S_{j}$ are the Mn spins on nearest neighbor sites, and hence
for any rotation away from  either of the two ground state spin
orientations, the free energy will increase. The transition across
the $T_{SR}$ boundary therefore is similar to a two level system
with an energy barrier associated with the intermediate state. The
field and temperature dependence of the peaks in the dielectric
constant shown in Fig. 8 should provide guidance for future
theoretical work to describe the behavior of the dielectric signal
as the $T_{SR}$ boundaries are crossed.

\subsection{\label{sec:level2} Significance of the doping studies}

As discussed above, the increase in Y (decrease in Ho)
concentration increases the region of the
$P\underline{6_{3}}c\underline{m}$ phase which is also the
symmetry of pure $YMnO_{3}$ below $T_{N}$. By reducing the Ho
concentration, it also takes higher B//c magnetic fields to induce
the P$\underline{6_{3}}c\underline{m}$ phase. Given the nature of
the sublattice spin structure, it is not clear how reducing the Ho
concentration increases the magnetic field at which the
$P\underline{6_{3}}c\underline{m}$ sublattice magnetic order
returns the $P\underline{6_{3}}$\underline{c}m configuration. At
x=0.8, the maximum field required to suppress ordering is
approximately 10~T, compared to the x=0 case where the maximum
field is about 4~T. This may be understood with the observation
that it is the Ho moments, partially oriented along the
c-direction, which cause the spin reorientation transition. The
addition of a magnetic field would presumably order the remaining
Ho moments, and with this internal field induce a transition
within the Mn layer to a paramagnetic state(through the
introduction of spin-flip transitions out of the basal plane).
Reducing the Ho spin concentration would reduce this internal
field, and thus the magnetic phase survives to higher applied
fields. Recent neutron scattering results have shown that the
anisotropy of the spin wave spectrum at low temperatures increases
greatly due to the presence of ordered Holmium, and drives the
spin reorientation transition.\cite{vajk} Ga (y=0.1) doping
reduces antiferromagnetic ordering of Mn sublattice and the
consequences are the increase of $T_{SR}$ and the decrease of
$T_{Ho}$. The slight decrease in Ho ordering temperature is one
more indication of mutual Ho-Mn interactions that drive the spin
reorientation and the rare earth ordering.

    The rare earth ordering
(i.e. $T_{Ho}$) is also suppressed with decreasing Ho
concentration, and this seems natural since the rare earth nearest
neighbor interactions will be reduced.  Due to the $\sim 4 K$
limit of our investigation, we did not explore the complicated
low-T phase diagram for $HoMnO_{3}$ reported by other research
groups.\cite{fiebig4,lorenz} We note that our spin reorientation
temperature for $HoMnO_{3}$ is higher than those reported in these
previous studies ($T_{SR}$ = 42~K vs. 33~K),where crystals were
flux grown. In the present case we used  single crystals, grown by
the floating zone method, and this may account for the higher
$T_{SR}$ values. We do however confirm the phase line near 5~K due
to the rare earth Ho ordering, and also the apparent widening of
the "INT" region at low temperatures.  This is within a region of
phase space where neutron experiments see a continuous change in
magnetic Bragg peak intensity(ie. as the spins change orientation
from one orientation to the next).\cite{fiebig1, vajk} The
hysteresis noted for $HoMnO_{3}$ below 5~K in fields between 0.5~T
and 2~T could be due to domain effects rather than true magnetic
transitions.

\subsection{\label{sec:level2} Significance of the magnetic field direction}

The Mn spins within the ab plane are severely frustrated, large
fields are needed to decouple them. Hence it is unlikely that in
the range of magnetic field studied here the Mn spins were
re-aligned solely due to the external field. Rather, it is most
likely the magnetic field interaction with the Ho spin system,
which is further coupled to the Mn spins, which drives the
transitions. Since the Ho magnetic sublattice spin orientations
are along the c-axis, a magnetic field applied in the ab plane
will involve various types of antiferromagnetic polarization
effects (spin flop, spin flip). Whereas for B//c the re-entrant
phases are still present, it is evident from Fig. 7 that for
B//ab, the re-entrance of the $P\underline{6_{3}}$\underline{c}m
phase at temperatures below the first $T_{SR}$ transition is
completely suppressed. Hence for B//ab, the interaction of the
magnetic field with the Ho system must be considerably stronger.
In Fig. 7 the phase line for x = 0 near 5~K is $T_{Ho}$, which is
suppressed for increasing field, and above 3~T, there is no
discernable transition at $T_{Ho}$. This gives a rough estimate of
the interactions between the Ho moments,which suggests an energy
scale of about 1.8~meV (using J=8). This is remarkably close to
the first crystal field level of Ho as determined by neutron
scattering(1.5(1)~meV), providing a mechanism for inducing spin
flip transitions for the Ising-like Ho moments.\cite{vajk}

\section{\label{sec:level1}SUMMARY AND CONCLUSIONS:\protect\\}

We have shown that by modifying the influence of the Ho sublattice
by either non-magnetic substitution for Ho, or with in-plane
magnetic fields, the mechanism that favors the
P$\underline{6_{3}}$\underline{c}m state is suppressed and the
$P\underline{6_{3}}c\underline{m}$ characteristic of the pure
$YMnO_{3}$ system emerges. To fully understand these effects, a
microscopic model is needed which includes a description of how
the magnetic field (applied in different directions) affects the
correlated Ho and Mn spin sublattices. Further experimental work
to characterize the nature of the spin systems vs. magnetic field
direction and alloy control, particularly neutron scattering and
second harmonic generation studies, would be very useful to better
understand these complex interacting spin systems.

\begin{acknowledgments}
   This research was sponsored by the National Nuclear Security Administration
   under the Stewardship Science Academic Alliances program through DOE
   Research Grant \# DE-FG03-03NA00066 (EJ), NSF Grant \#'s DMR0203532 and DMR0602859 (JSB),
   and the NHMFL is supported by contractual agreement between
   the National Science Foundation through NSF Grant
   \# DMR0449569 and the State of Florida.
\end{acknowledgments}

\begin{figure}[ht]
\includegraphics[width=8cm]{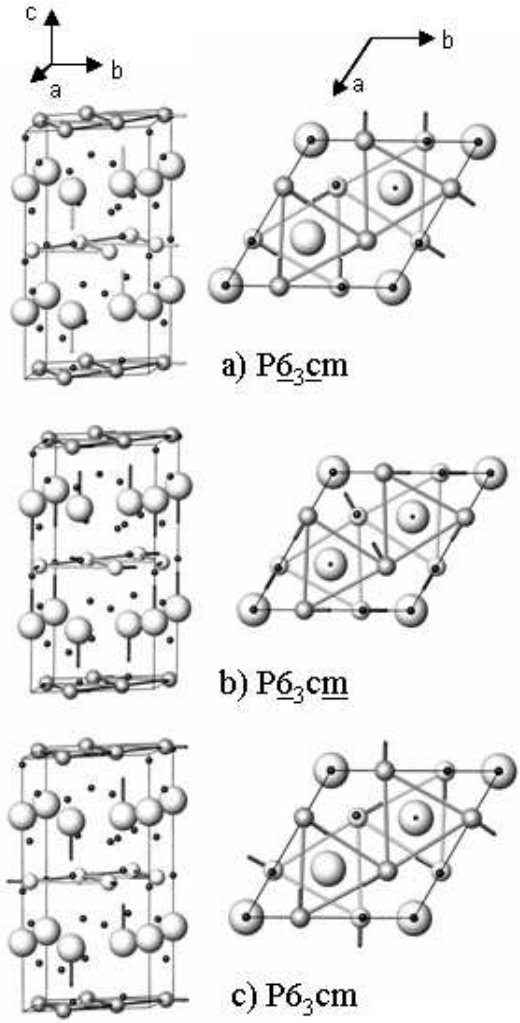}
\caption{Spin configurations and symmetries of hexagonal
$HoMnO_{3}$ below $T_N$ (after Ref.\cite{lottermoser}) for the
side(left)and top(right)views of hexagonal unit cell. a)
$P\underline{6_{3}c}m, T_{SR} < T < T_{N}$, b)
$P\underline{6_{3}}c\underline{m}, T_{Ho}<T< T_{SR}$; c)
$P6_{3}cm,  T< T_{Ho}$. Legend: thinner lines - boundaries of the
hexagonal unit cell;  thicker lines in the ab plane - triangular
antiferromagnetic configuration; large spheres - $Ho^{3+}$; small
spheres - $Mn^{3+}$ where central ab plane ions are lighter color;
black dots - $O^{2-}$; vertical arrows - Ho magnetic moments;
arrows in ab plane - Mn spins} \label{fig1}
\end{figure}

\begin{figure}[ht]
\includegraphics[width=8cm]{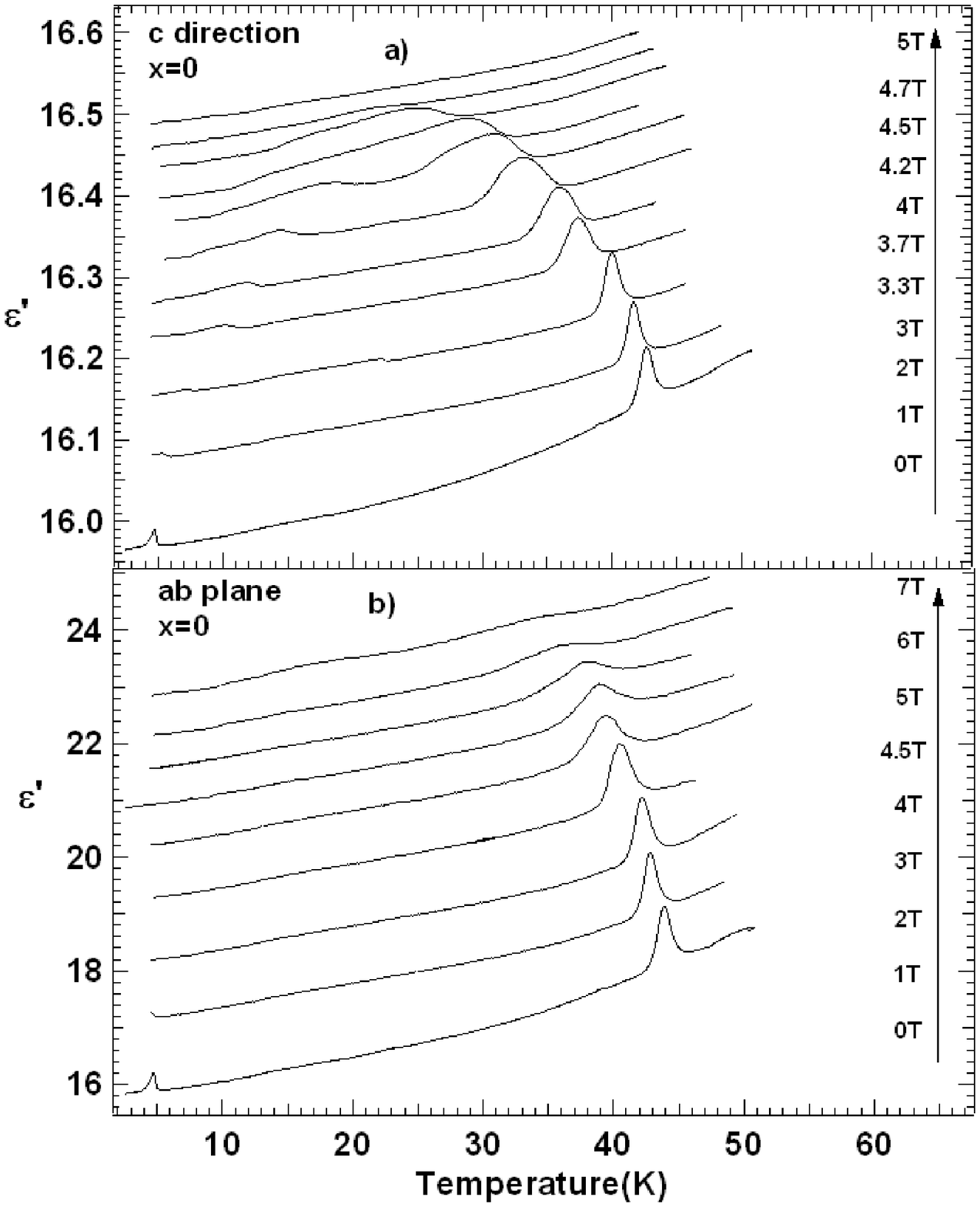}
\caption{Temperature dependence of real part of dielectric
constant for $Ho_{1-x}Mn_{x}O_{3}$, (x = 0) for different magnetic
fields. For $B > 0$, the curves are shifted upwards by arbitrary
amounts. a) B//c. Here $T_{SR}$ and $T_{Ho}$ are indicated for B =
0. b) B//ab.} \label{fig2}
\end{figure}

\begin{figure}[ht]
\includegraphics[width=8cm]{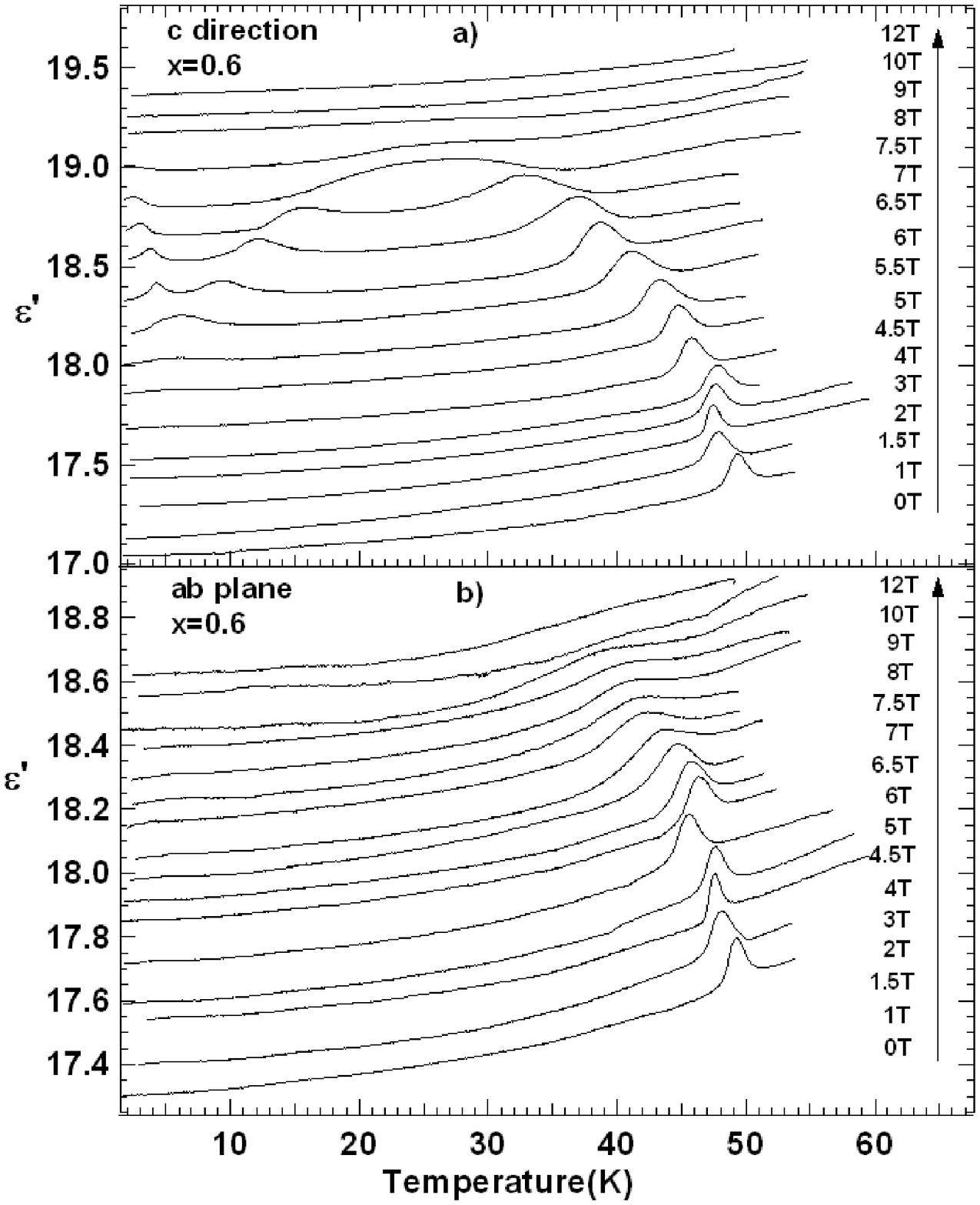}
\caption{Temperature dependence of real part of dielectric
constant for $Ho_{1-x}Mn_{x}O_{3}$, (x = 0.6 ) for different
magnetic fields. For $B > 0$, the curves are shifted upwards by
arbitrary amounts. a) B//c. b) B//ab.} \label{fig3}
\end{figure}

\begin{figure}[ht]
\includegraphics[width=8cm]{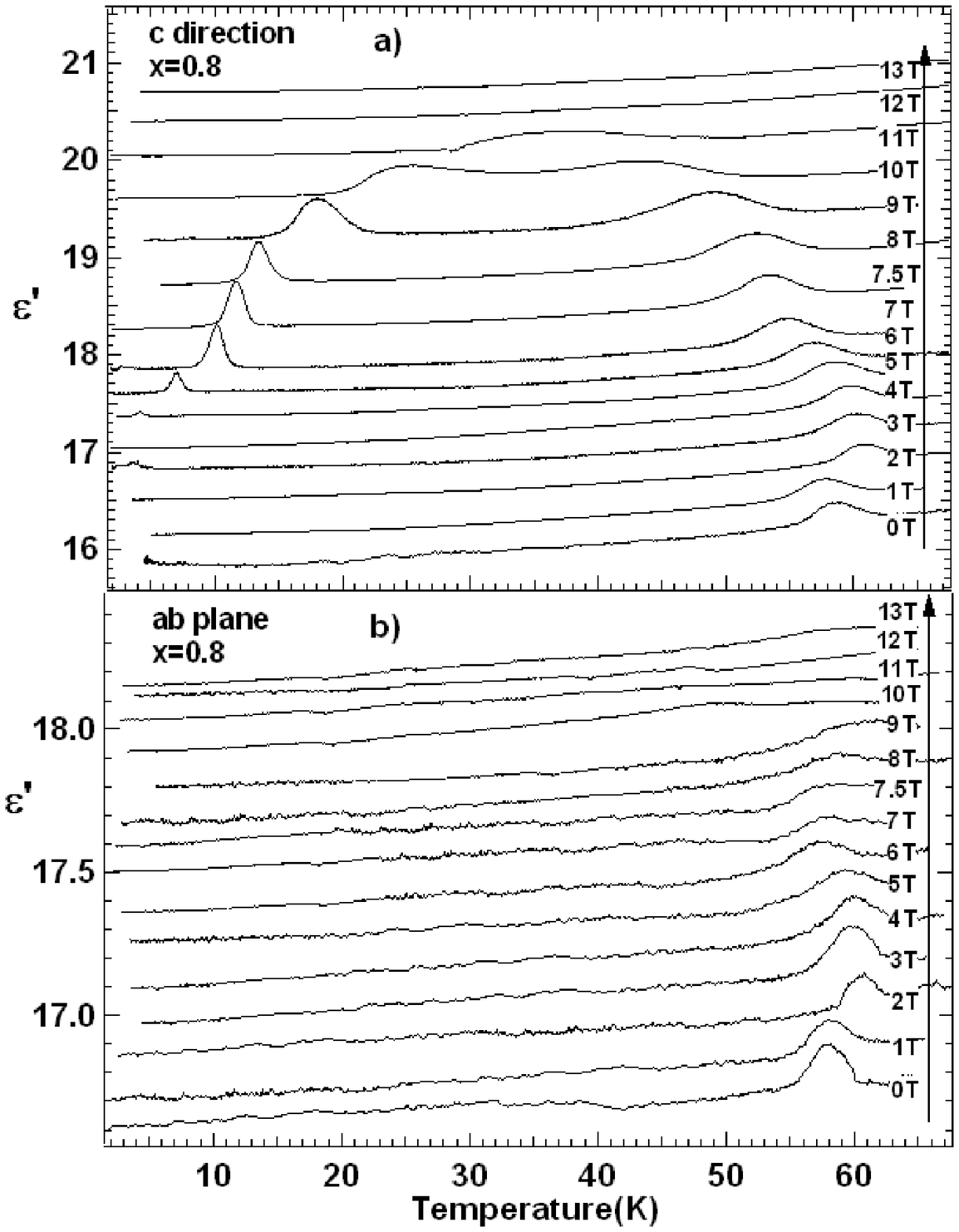}
\caption{Temperature dependence of real part of dielectric
constant for $Ho_{1-x}Mn_{x}O_{3}$, (x = 0.8 ) for different
magnetic fields. For $B > 0$, the curves are shifted upwards by
arbitrary amounts. a) B//c. b) B//ab.} \label{fig4}
\end{figure}

\begin{figure}[ht]
\includegraphics[width=8cm]{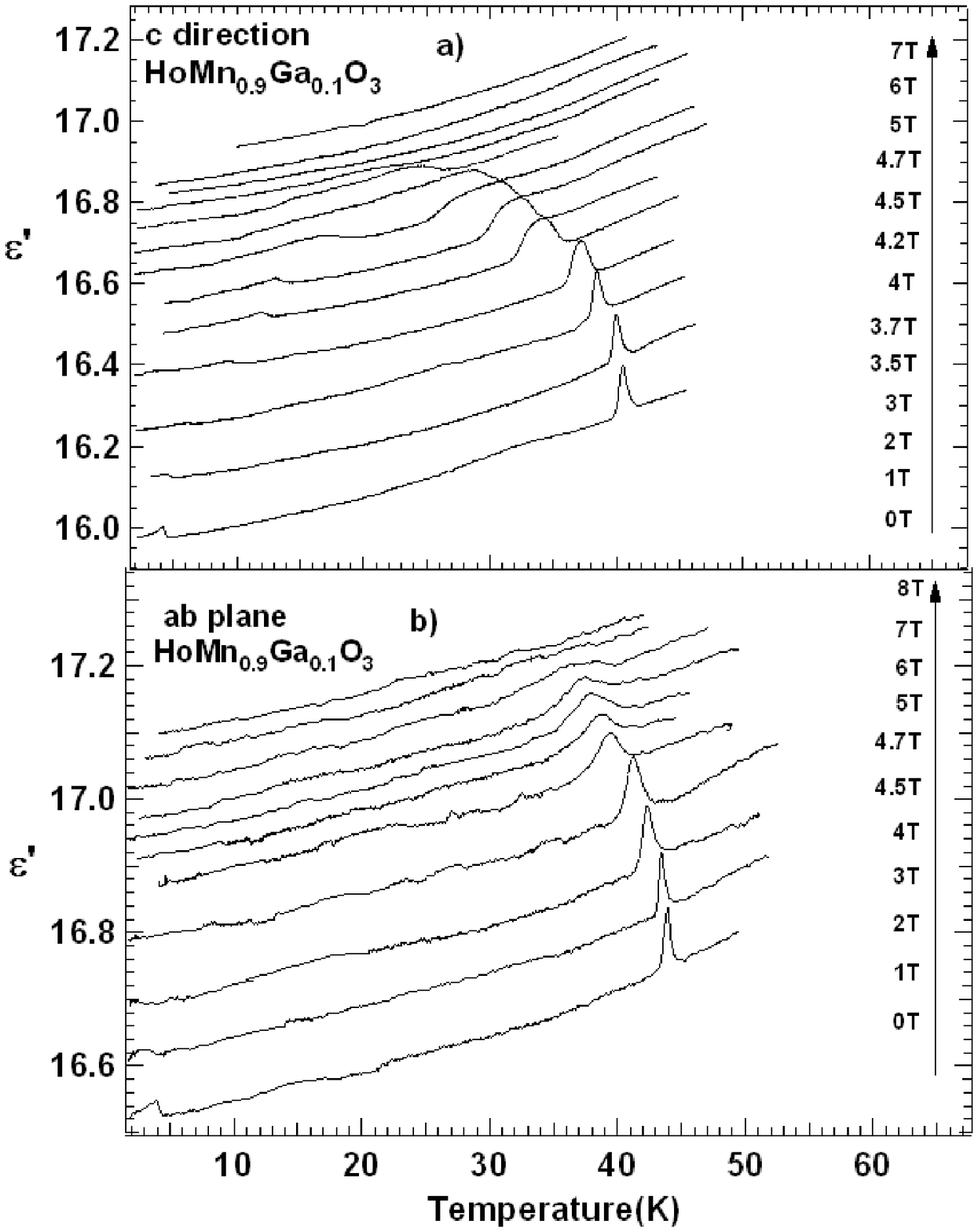}
\caption{Temperature dependence of real part of dielectric
constant for $HoMn_{1-y}Ga_{y}O_{3}$, (y = 0.1) for different magnetic
fields. For $B > 0$, the curves are shifted upwards by arbitrary
amounts. a) B//c. b) B//ab.} \label{fig5}
\end{figure}

\begin{figure}[ht]
\includegraphics[width=8cm]{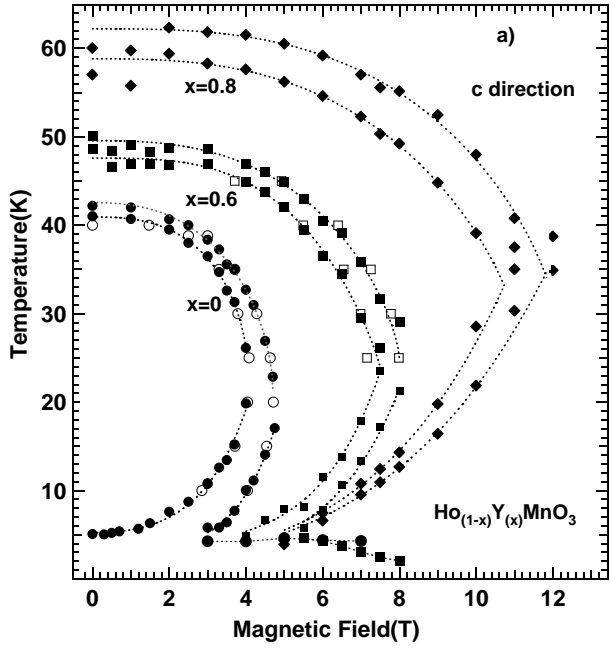}
\caption{Magnetic field-temperature phase diagrams for
$Ho_{1-x}Y_{x}MnO_{3}$ $(x=0; 0.6; 0.8)$ for magnetic field
applied in c direction. Solid and open symbols represent
temperature and field sweeps respectively. Circles, $x=0$;
squares, $x=0.6$; diamonds, $x=0.8$. The intermediate phase
boundary widths correspond to the dielectric constant peak
half-widths. The phase diagram for $HoMn_{0.9}Ga_{0.1}O_{3}$ for
B//c (not shown) is indiscernible from that for pure $HoMnO_{3}$.}
\label{fig7}
\end{figure}

\begin{figure}[ht]
\includegraphics[width=8cm]{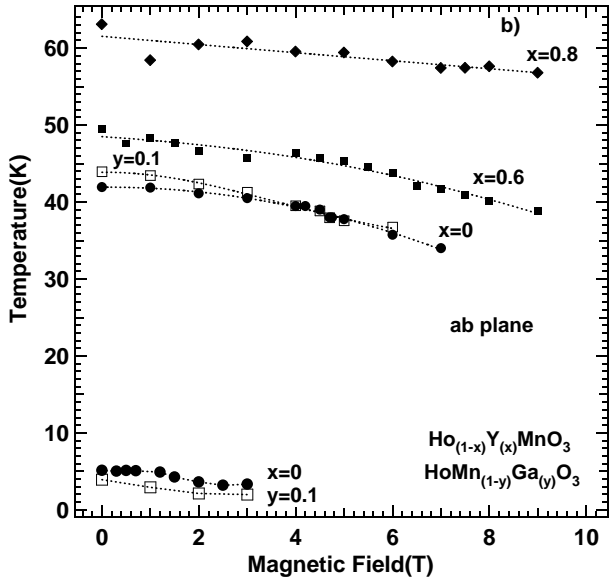}
\caption{Magnetic field-temperature phase diagrams for
$Ho_{1-x}Y_{x}MnO_{3}$ $(x=0; 0.6; 0.8)$ and
$HoMn_{1-y}Ga_{y}O_{3}$ (y = 0.1) for magnetic field applied in
ab-plane direction. All data is from temperature sweeps, and only
the dielectric peak locations are indicated. Circles, $x=0$;
squares, $x=0.6$; diamonds, $x=0.8$; open squares, $y =0.1$. }
\label{fig8}
\end{figure}

\begin{figure}[ht]
\includegraphics[width=8cm]{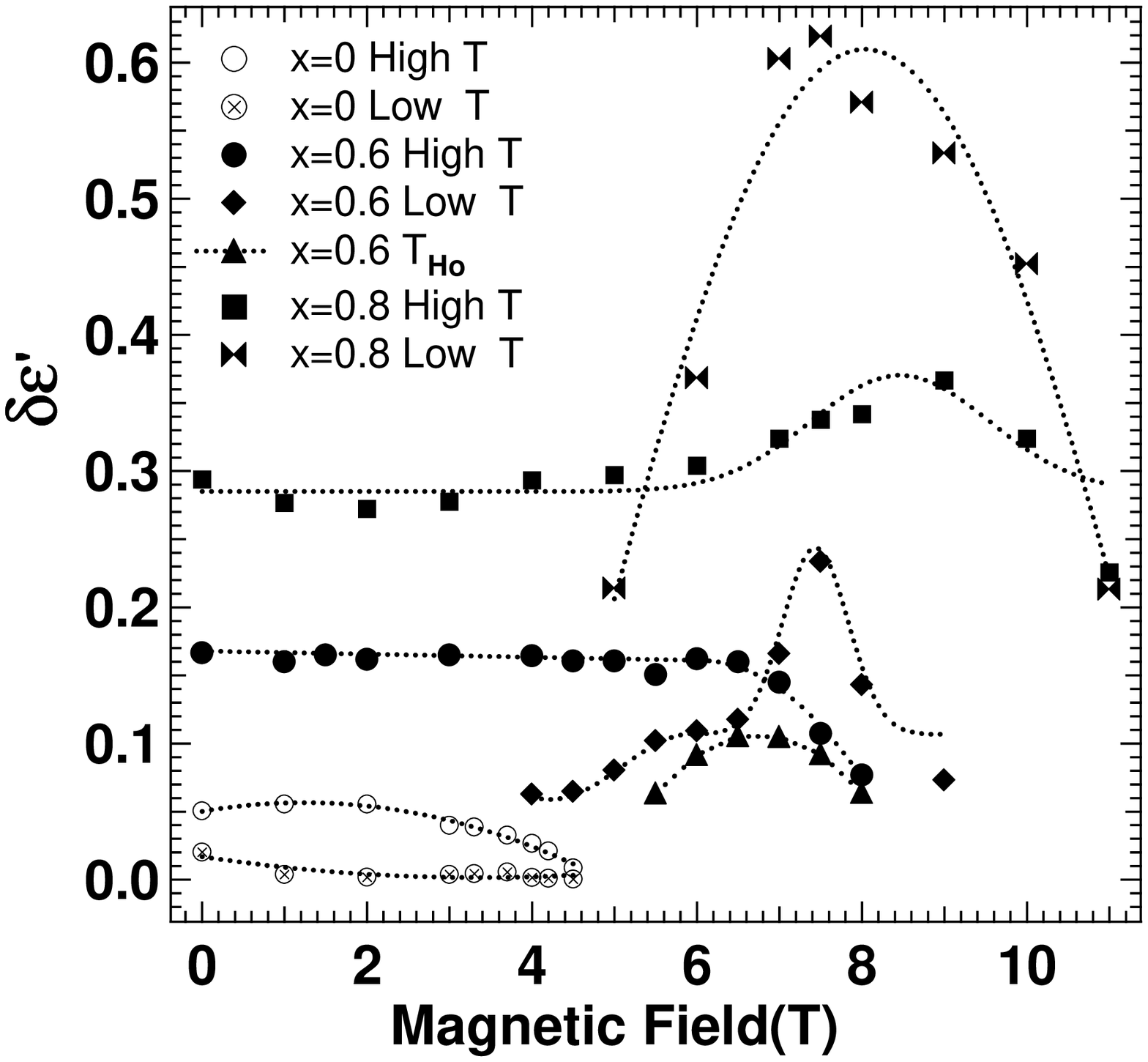}
\caption{Field dependence of the dielectric peak heights
$\delta\epsilon'$ at the $T_{SR} $ and $T_{Ho}$ transitions for the
$Ho_{1-x}Y_{x}MnO_{3}$ compounds for B//c. Note that the peak
locations also depend on temperature, as indicated in the legend
as "High" and "Low" for the upper and lower temperature branches
of the $T_{SR}$ phase boundaries respectively (similarly for
$T_{Ho}$).} \label{fig9}
\end{figure}


\begin{thebibliography}{}
\bibitem{fiebig1}See for example M. Fiebig, T. Lottermoser, D.
Frohlich, A. V. Goitsev, and R. V. Pisarev, Nature 419, 818 (2002).
\bibitem{bertaut}F. Bertaut, P. Fang, and P. Forrat, Comptes Rendus
Hebdomadaires Des Seances de L'Academie des Sciences 256, 1958
(1963).
\bibitem{huang}Z. J. Huang, Y. Cao, Y. Y. Sun, Y. Y. Xue, and C. W.
Chu, Phys. Rev. B. 56, 2623 (1997).
\bibitem{hill}N. A. Hill, J.  Phys. Chem. B 104, 6694 (2000).
\bibitem{coeure}P. Coeuré, P. Guinet, J. C. Peuzin, G. Buisson, and E. F. Bertaut, Proceedings of the International Meeting on Ferroelectricity, edited by V. Dvorák, A. Fousková, and P. Glogar (Institue of Physics of the Czechoslovak Academy of Science, Prague 1966) Vol. 1, pp. 332-340.
\bibitem{fiebig2}M. Fiebig, Th. Lottermoser, and R. V. Pisarev, J. Appl. Phys. 93, 8194 (2003).
\bibitem{vajk}O. P. Vajk, M. Kenzelmann, J. W. Lynn, S. B. Kim, S. W. Cheong, Phys. Rev. Lett. 94, 087601 (2005).
\bibitem{lorenz}B. Lorenz, F. Yen, M. M. Gospodinov, and C. W. Chu, Phys. Rev. B. 71, 014438 (2005).
\bibitem{munoz}A. Muñoz, J. A. Alonso, M. J. Martínez-Lope, M. T. Casáis, J. L. Martínez, and M. T. Fernández-Díaz, Chem. Mater. 13, 1497 (2001).
\bibitem{lonkai}Th. Lonkai, D. Hohlwein, J. Ihringer, and W. Prandl, Appl. Phys. A: Mater. Sci. Process. 74, 843 (2002).
\bibitem{fiebig3}M. Fiebig, D. Fr\"{o}hlich, K. Kohn, St. Leute, Th. Lottermoser, V. V. Pavlov, and R.
V. Pisarev, Phys. Rev. Lett. 84, 5620 (2000).
\bibitem{lottermoser}Th. Lottermoser, Th. Lonkai, U. Amann, D. Hohlwein, J. Ihringer, and M. Fiebig, Nature 430, 541 (2004).
\bibitem{fiebig4}M. Fiebig, C. Degenhardt, and R. V. Pisarev, J. Appl. Phys. 91, 8867 (2002)
\bibitem{yen}F. Yen, C. R. dela Cruz, B. Lorenz, Y. Y. Sun, Y. Q. Wang, M. M. Gospodinov, and C. W. Chu, Phys. Rev. B, 71, 180407(R) (2005).
\bibitem{lee}Seongsu Lee, A. Pirogov, Jung Hoon Han, J. -G. Park, A. Hoshikawa, and T. Kamiyama, Phys. Rev. B, 71, 180413(R) (2005).
\bibitem{cruz}C. R. dela Cruz, F. Yen, B. Lorenz, Y. Q. Wang, Y. Y. Sun, M. M. Gospodinov, and C. W. Chu, Phys. Rev. B, 71, 060407(R) (2005).
\bibitem{litvinlorentPRL}B. Lorenz, A. P. Litvinchuk, M. M. Gospodinov, C. W.  Chu, Phys. Rev. Lett. 92, 087204 (2004).
\bibitem{katsufuji}T. Katsufuji, S. Mori, M. Masaki, Y. Moritomo, N. Yamamoto, H. Takagi, Phys. Rev. B 64, 104419 (2001).
\bibitem{hdzhou}H. D. Zhou, J. C. Denyszyn, J. B. Goodenough, Phys. Rev. B 72, 224401 (2005)
\bibitem{sharma} P. A. Sharma, J. S. Ahn, N. Hur, S. Park, Sung Baek Kim, Seongsu Lee, J.-G. Park, S. Guha, and S-W. Cheong, Phys. Rev. Lett. 93, 177202 (2004)
\bibitem{zhong}C. G. Zhong and Q. Jiang, J. Phys. Condens. Matter 14, 8605(2002)

\end{thebibliography}
\end{document}